\documentclass{article}

\usepackage{amssymb}
\usepackage{amsmath}
\usepackage{stmaryrd}
\usepackage[margin=1in]{geometry}
\usepackage{graphics}
\usepackage{tikz}
\usepackage{epsfig}
\usepackage{multirow}
\usepackage{graphicx}
\usepackage[center]{caption}
\usepackage{subcaption}
\usepackage{cancel}
\usepackage{upquote}
\usepackage[hidelinks]{hyperref}
\usepackage{graphicx} 
\newcommand{\sam}{S_{\text{a-m}}}

	\title{Two-parameter superposable S-curves}
\author{Vijay Prakash S\\
		\small Independent Researcher, Alappuzha, Kerala, India.\\
	\small \tt{prakash.vijay.s@gmail.com}}
\date{}
\begin{document}
	\maketitle
	\begin{abstract}
\noindent
Straight line equation $y=mx$ with slope $m$, when singularly perturbed as $ay^3+y=mx$ with a positive parameter $a$, results in S-shaped curves or S-curves on a real plane. As $a\rightarrow 0$, we get back $y=mx$ which is a cumulative distribution function of a continuous uniform distribution that describes the occurrence of every event in an interval to be equally probable. As $a\rightarrow\infty$, the derivative of $y$ has finite support only at $y=0$ resembling a degenerate distribution. Based on these arguments, in this work, we propose that these S-curves can represent maximum entropy uniform distribution to a zero entropy single value. We also argue that these S-curves are superposable as they are only parametrically nonlinear but fundamentally linear. So far, the superposed forms have been used to capture the patterns of natural systems such as nonlinear dynamics of biological growth and kinetics of enzyme reactions. 
Here, we attempt to use the S-curve and its superposed form as statistical models. We fit the models on a classical dataset containing flower measurements of iris plants and analyze their usefulness in pattern recognition. Based on these models, we claim that any non-uniform pattern can be represented as a singular perturbation to uniform distribution. However, our parametric estimation procedure have some limitations such as sensitivity to initial conditions depending on the data at hand.
	\end{abstract}
	
	\section{Introduction}
	\label{sec_intro}
Usually represented as a modified form of exponential function, sigmoidal or S-shaped pattern is observed in various scientific  domains \cite{appl_S}. The sigmoidal function is given by
	\begin{equation}
S_{\text{exp}}(x) = \frac{1}{1+\text{e}^{-x}}.
\label{eq_sigm}
	\end{equation}
Superposition of the above S-curve by Cybenko \cite{cybenko_89} has led to the development of universal function approximation by Hornik et al \cite{hornik_89}. Other similar modified forms of exponential function include the logistic function and its extensions \cite{analy_growth, log_sigm, forest_gr, turner_76}.

Typically, S-curves are integrals of bell-shaped curves. De Moivre was the first to modify the exponential function to represent errors in measurements as bell-shaped curves \cite{stahl}. De Moivre's modified form was later parametrized by Gauss now known as the normal distribution function given by
\begin{equation}
\tilde{y}=\frac{1}{\sqrt{2\pi}\sigma}\exp^{-\frac{1}{2}\left(\frac{x-\mu}{\sigma}\right)^2}
\label{eq_bell_sigm}
\end{equation}
which are two-parameter family of curves. However, once the mean $\mu$ is located the height and tails of the bell-curve is controlled by just one parameter $\sigma$. The above bell-curve $\tilde{y}$ is the derivative of Gauss error function which is also an S-curve. $\tilde{y}\rightarrow0$ for both $\sigma\rightarrow0$ and $\sigma\rightarrow\infty$. So, $\tilde{y}$ is applicable only for distributions with finite variance\cite{taleb_book}.

S-curves can also be obtained algebraically. A singular perturbation to the straight line equation $y=mx$ with slope $m$ also results in S-curves given by \cite{shruti_vij,shruti_Ssum}
\begin{equation}
ay^3+y=mx,
\label{eq1}
\end{equation}
where $a$ is a positive parameter. The generalized form of Eqn. (\ref{eq1}) is given by
\begin{equation}
y-y_c = \frac{m(x-x_c)}{1+a(y-y_c)^2},
\label{eq_gen}
\end{equation}
where $(x_c,y_c)$ is the point of inflection. Corresponding bell-curves are given by
\begin{equation}
\frac{dy}{dx} = \frac{m}{1+a(y-y_c)^2}.
\label{eq_gen_der}
\end{equation}
At the point of inflection $y=y_c$, $\dfrac{dy}{dx}=m$. Unlike Eqn. (\ref{eq_bell_sigm}), a bell-curve  described with the above equation has two parameters $a$ and $m$. Its height is described by $m$ and both $a$ and $m$ describe the tails. 
The real solution of Eqn. (\ref{eq_gen}) is given by
\begin{equation}
y(a,m,x_c,y_c) = S_1(a,m,x-x_c)+S_2(a,m,x-x_c)+y_c.
\label{eq_gen_y}
\end{equation}
With $\hat{t}=-\left(\dfrac{27m(x-x_c)}{2a}\right)+\sqrt{\left(\dfrac{27m(x-x_c)}{2a}\right)^2+\dfrac{27}{a^3}}$, $S_1= \dfrac{-1}{3}\hat{t}^{{1}/{3}}$ and $S_2= \dfrac{1}{a}\hat{t}^{{-1}/{3}}$. $S_1$ dominates the lower end of the S-curves whereas $S_2$ dominates the other part.

\begin{figure}[ht!]
	\centering
	\begin{subfigure}[t]{0.45\textwidth}
		\includegraphics[width=\textwidth]{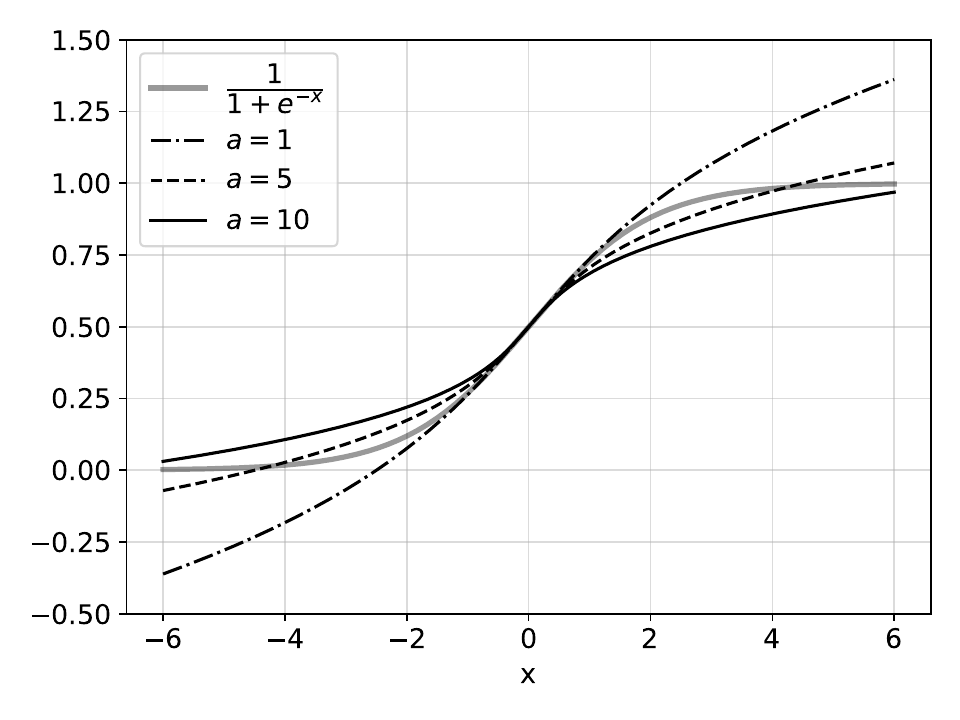}
		\caption{Comparison of $(a,m)-$ S-curves with $S_{\text{exp}}$ with $m=0.25$ and $(x_c,y_c)$ as (0,0.5) for various $a$.}
		\label{fig1a}
	\end{subfigure}
	\begin{subfigure}[t]{0.45\textwidth}
		\includegraphics[width=\textwidth]{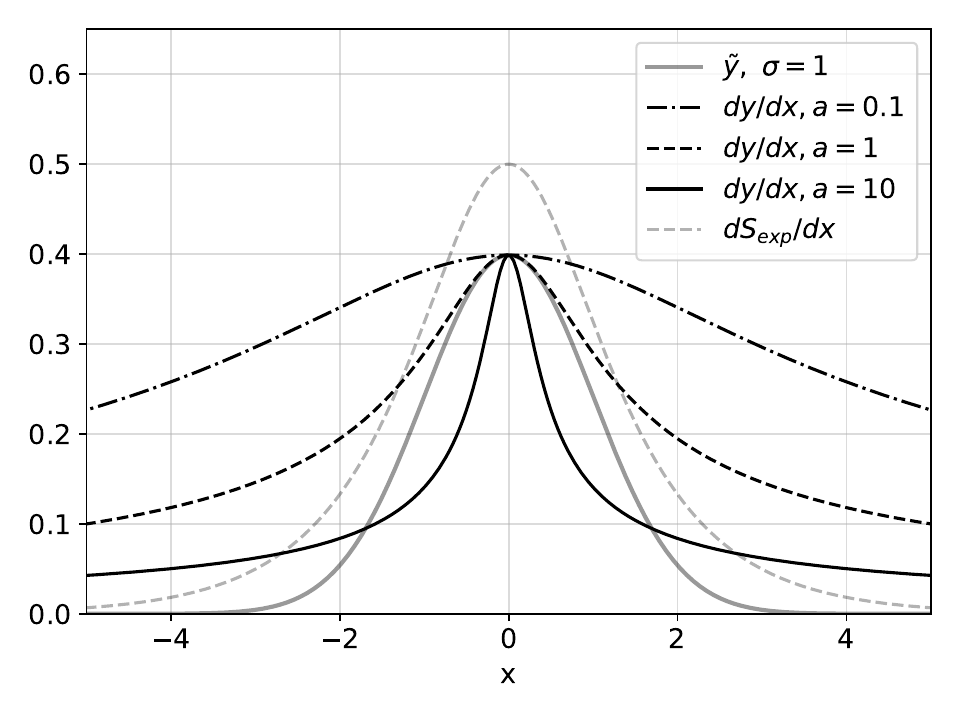}
		\caption{$m=1/\sqrt{2\pi}$. As $a\rightarrow0$ the distribution becomes uniform and as $a\rightarrow\infty$ the distribution resembles a delta function.}
		\label{fig1b}
	\end{subfigure}
	\caption{S- and bell- shaped curves of modified exponential function and modified (singularly perturbed) straight lines.}
	\label{fig1}
\end{figure}

In Fig. \ref{fig1}, Eqn. (\ref{eq_gen_y}) referred to as $S_{\text{a-m}}$ curves and bell- curves of Eqn. (\ref{eq_gen_der}) are compared with those of modified exponential function forms (Eqn. (\ref{eq_sigm}) and (\ref{eq_bell_sigm})). 
Originally introduced as an adjustment to $y-$ axis in order to work with bounded dependent variable even for large values of the independent variable, Eqn. (\ref{eq_gen}) has been used to classify the images of the fashion-MNIST dataset \cite{1st_paper} and to model biological growth \cite{shruti_vij}.

In the case of image classification, the parameter $a$ acts as a regularizer and offers adaptive learning rate in a gradient descent based estimation. In \cite{1st_paper}, each of $28\times28$ pixel value is represented as a component of $\textbf{x}$ and the input representation is a linear combination given by
\begin{eqnarray}
w_0+\sum_{i=1}^{n=784} w_ix_i = w^T\textbf{x}
\label{eqwx}
\end{eqnarray}
where $w_i$'s are weights. This input is activated with a two-parameter $S_{\text{a-m}}$ curve for each category
\begin{equation}
ay^3+y=m(w^T\textbf{x}).
\label{eq_gen_im}
\end{equation}
Thus instead of fixed hyperparametric values, we have parametric tuning of regularization and learning rate. This leads to quick convergence of logistic regression without Tikhonov\cite{tikhonov_43} or lasso \cite{tibshirani_96} regularization of weights with their learning rate set to unity. With Eqn. (\ref{eq_gen_im}), the image classification accuracy is around 84\% which means that 84\% of Fashion-MNIST images fall under a symmetric distribution described by Eqn. (\ref{eq_gen_der}) when the images are represented as a linear combination of normalized pixel values.
In the case of biological growth, $a$ acts as a restriction parameter that introduces nonlinear restricting influences on linear growth with growth rate $m$. The model (Eqn. \ref{eq_gen}) fits the symmetric portions around the point of maximum growth, which is $(x_c,y_c)$. 

In Fig. \ref{fig1}, it can be seen that the $S_{\text{a-m}}$ curves do not offer the same curvature as $S_{\text{exp}}$ of Eqn. (\ref{eq_sigm}) or $\tilde{y}$ (Eqn. (\ref{eq_bell_sigm})). Of course, the nonlinearity of exponential functions is unmatched with a simple parametric algebraic expression (Eqn. (\ref{eq_gen})) alone. So the natural choice would be to introduce more nonlinear terms in Eqn. (\ref{eq_gen}) which may lead to a power series in $y$ and more parameters as coefficients of high powers of $y$. Instead of such a complicated procedure, in the following section, we simply superpose various $S_{\text{a-m}}$ curves with different origins (or points of inflections) and fit the superposition on the $S_{\text{exp}}$ curve and the Gauss error function and obtain $\tilde{y}$ as its derivative.

\section{Superposition of $\sam$ curves}
\begin{figure}[ht!]
	\centering
	\begin{subfigure}[t]{0.45\textwidth}
		\includegraphics[width=\textwidth]{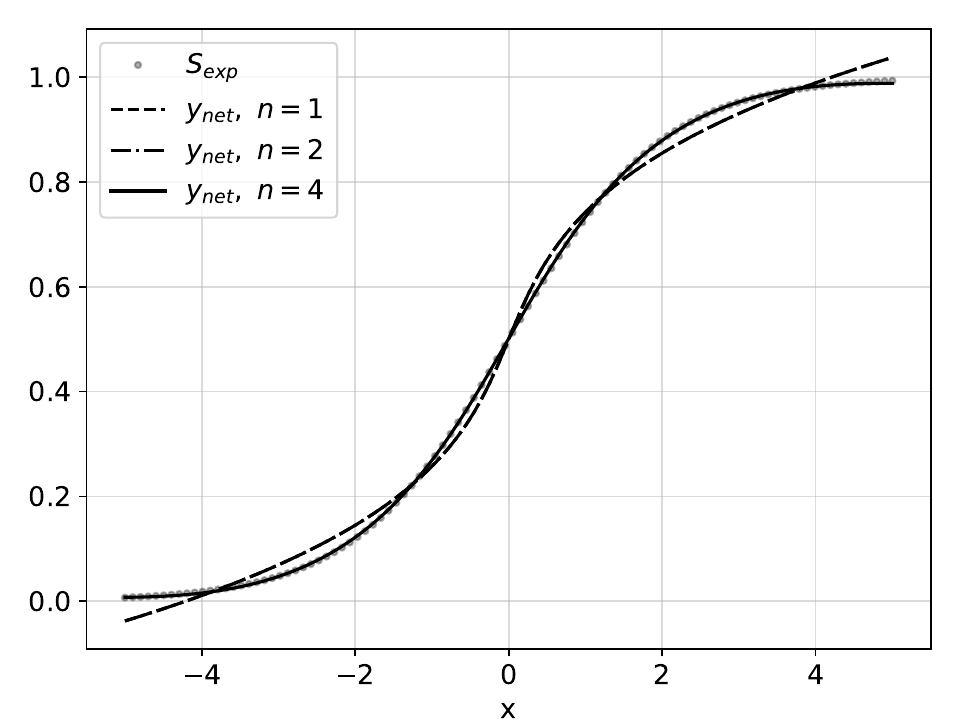}
		\caption{Fitting superposed $S_{\text{a-m}}$ curves on $S_{\text{exp}}$. Fits for $n=2$ and $n=4$ lie on top of each other.}
		\label{fig2a}
	\end{subfigure}
	\begin{subfigure}[t]{0.45\textwidth}
		\includegraphics[width=\textwidth]{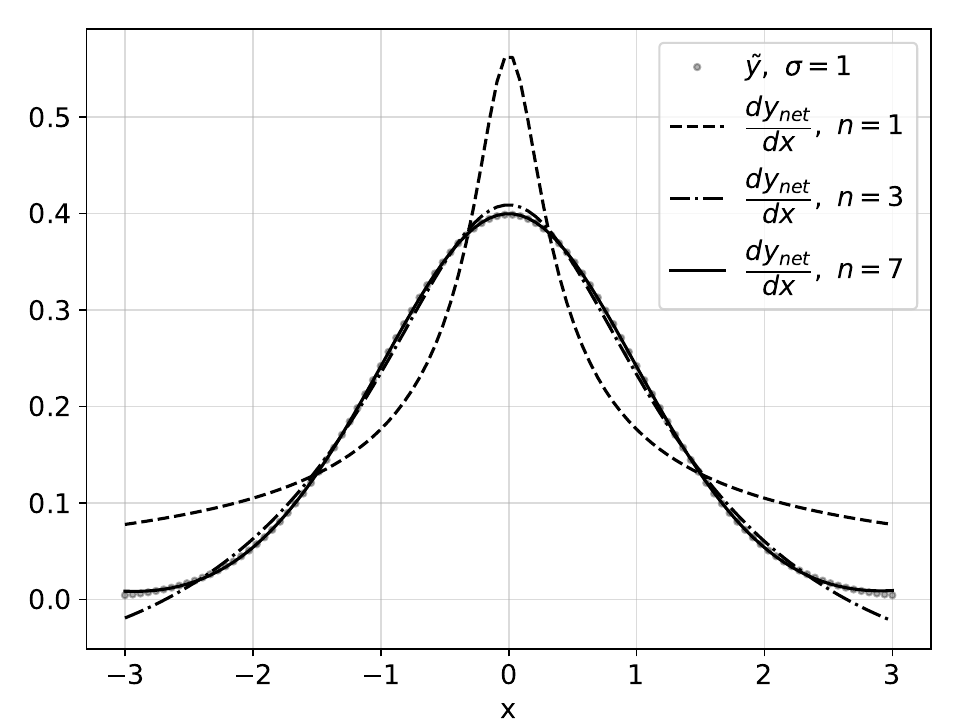}
		\caption{Derivative of superposed  $S_{\text{a-m}}$ curves fitted on the Gauss error function.}
		\label{fig2b}
	\end{subfigure}
	\caption{Fitting superposed $S_{\text{a-m}}$ curves on S- and bell- shaped curves obtained by modifying the exponential function. We get better fits as more $S_{\text{a-m}}$ curves are added to the superposition.}
	\label{fig2}
\end{figure}

From Eqn. (\ref{eq_gen}), we make the following observations and decide upon the superposed form.
\begin{enumerate}
	\item Within  data, there can be large variations in slope values yet a system under study is bounded and remains measurable. Even with large variations in $m$, $a$ parameterizes the nonlinear adjustment to $y$ that keeps $y$ bounded. So, $a$ can be considered as an adjustment parameter or a nonlinear parameter as it turns on nonlinearities in Eqn. (\ref{eq_gen}).
	
	\item A slowly varying nonlinear curve such as $S_{\text{a-m}}$ has only one point of inflection. When there are large variations that requires special tools such as the exponential function, it is assumed that multiple points of inflection exist causing such variations. 
	
	\item Since $m$ is the value of slope at the inflection point $(x_c,y_c)$, in order to fit different variations in $x$, we superpose $S_\text{a-m}$ curves of different slopes at different inflection points under a common $y-$ axis adjustment by fixing $a$.
\end{enumerate}
Therefore, $S_{\text{a-m}}$ curves of different origins or inflection points are superposed in the following way
	\begin{eqnarray}
		\nonumber y_{\text{net}} &=& \qquad\sum_{i=1}^n p_iy(a,m_i,x_{ci},y_{ci})\\
		\Rightarrow y_{\text{net}} &=& \qquad\sum_{i=1}^n p_i\left(\frac{m_i(x-x_{ci})}{1+a(y_i-y_{ci})^2}+y_{ci}\right).
		\label{eqn_sup}
	\end{eqnarray}
	where $p_i$'s are weights, $m_i$ is the $i^{\text{th}}$ slope with $x_{ci}$ and $y_{ci}$ as the origin coordinates or points of inflection. From (\ref{eq_gen_y}), $y_i\equiv y(a,m_i,x_{ci},y_{ci})=S_1(a,m_i,x-x_{ci})+S_2(a,m_i,x-x_{ci})+y_{ci}$.
The inflection points are chosen from the given data. This is done by choosing the midpoint of data points with the highest absolute slope values. This is implemented with python programming tool as follows:
\begin{verbatim}
#ams is the number of origins 
grads=(ydata[1:]-ydata[:-1])/(xdata[1:]-xdata[:-1])
mxpts = np.argsort(np.abs(grads))[-ams:][::-1]
xcs = np.empty(ams);ycs = np.empty(ams)
for i in np.arange(ams):
  xcs[i] = (x[mxpts[i]]+x[mxpts[i]+1])/2
  ycs[i] = (y[mxpts[i]]+y[mxpts[i]+1])/2
\end{verbatim}
We consider absolute slopes because $m$ is allowed to take positive or negative values. We fit the linear combination on $S_{\text{exp}}$ and $\tilde{y}$ curves in Fig. \ref{fig2a} and \ref{fig2b}, respectively. The above superposition was first used to fit a bacterial growth dataset \cite{shruti_vij2}. Here, we fit the superposition on the $S_{\text{exp}}$ and $\tilde{y}$ functions in Fig. \ref{fig2a} and \ref{fig2b}, respectively.
To explain the fits in Fig. \ref{fig2}, consider two $S_{\text{a-m}}$ curves that are superposed, for $n=2$ we get
\begin{eqnarray}
\nonumber y_{\text{net}}&=&p_1\frac{m_1(x-x_1)}{1+a(y_1-y_{c1})^2}+p_2\frac{m_2(x-x_2)}{1+a(y_2-y_{c2})^2}\\
&=&\frac{p_1m_1(x-x_1)(1+a(y_2-y_{c2})^2)+p_2m_2(x-x_2)(1+a(y_1-y_{c1})^2)}{(1+a(y_1-y_{c1})^2)(1+a(y_2-y_{c2})^2)}.
\end{eqnarray}
It can be seen that as we superpose more $y_i$'s there are higher powers of $y_i$'s in the $y_{\text{net}}$ expression. Unlike higher order polynomials in $x$, the parametric estimation with $S_{\text{a-m}}$ curves is bounded and we do not directly find the coefficient values of higher order terms. This way high orders with $n=11$ has been fitted on a growth data of a human male \cite{montbeil}. 

For $n=2$, there are 5 parameters to be estimated they are $a$,$p_1$,$m_1$,$p_2$ and $m_2$. Superposition of 3 $S_{\text{a-m}}$ curves leads to a 7-parametric fit. With $n=4$, we fit a 9-parametric superposition and so on. 
\subsection{Measures from superposition}
\begin{figure}[ht!]
	\centering
	\begin{subfigure}[t]{0.45\textwidth}
		\includegraphics[width=\textwidth]{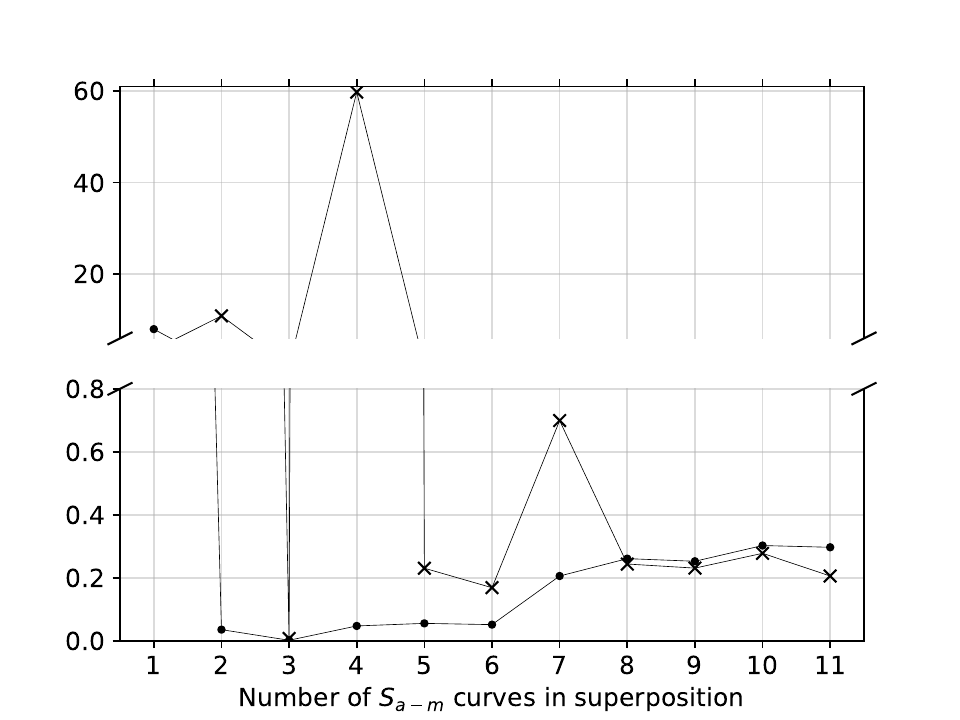}
		\caption{Estimation of $a$ stabilizes to some extent as $n$ increases.}
		\label{fig3a}
	\end{subfigure}
	\begin{subfigure}[t]{0.45\textwidth}
		\includegraphics[width=\textwidth]{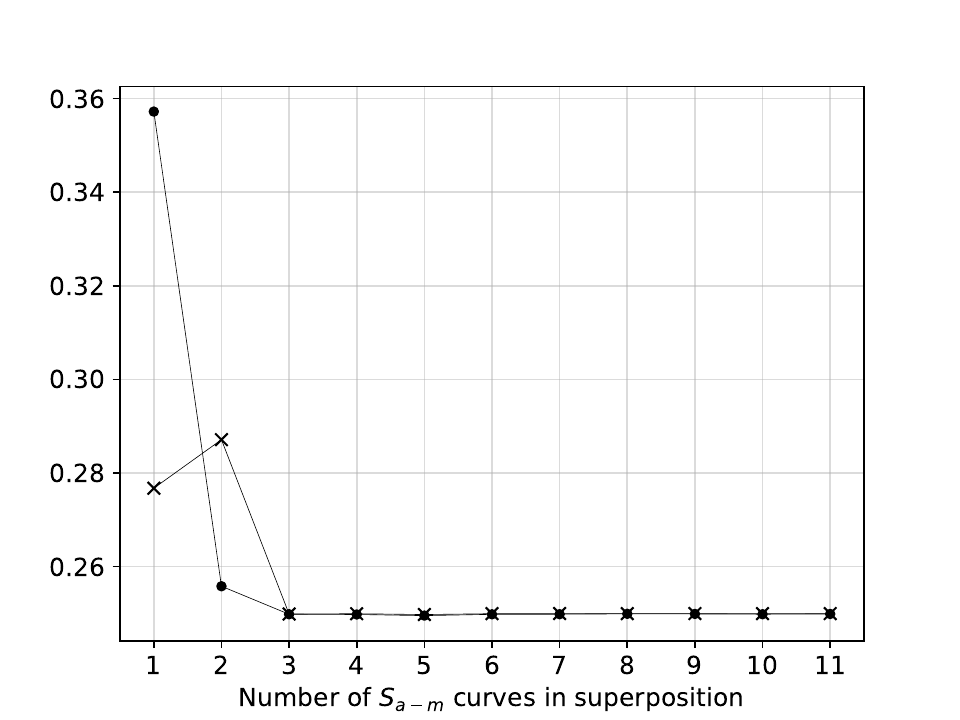}
		\caption{The maximum slope value $m$ converge to a more accurate estimation as $n$ increases.}
		\label{fig3b}
	\end{subfigure}
	\caption{Variation of parameters $a$ and $m$ based on the fits on $S_{\text{exp}}$ for two different intervals $x\in[-3,3]$ (shown as `$\times$') and $x\in[-5,5]$ (shown as `$\bullet$') with $n$ as the $x-$ axis.}
	\label{fig3}
\end{figure}

\begin{figure}[ht!]
	\centering
	\begin{subfigure}[t]{0.45\textwidth}
		\includegraphics[width=\textwidth]{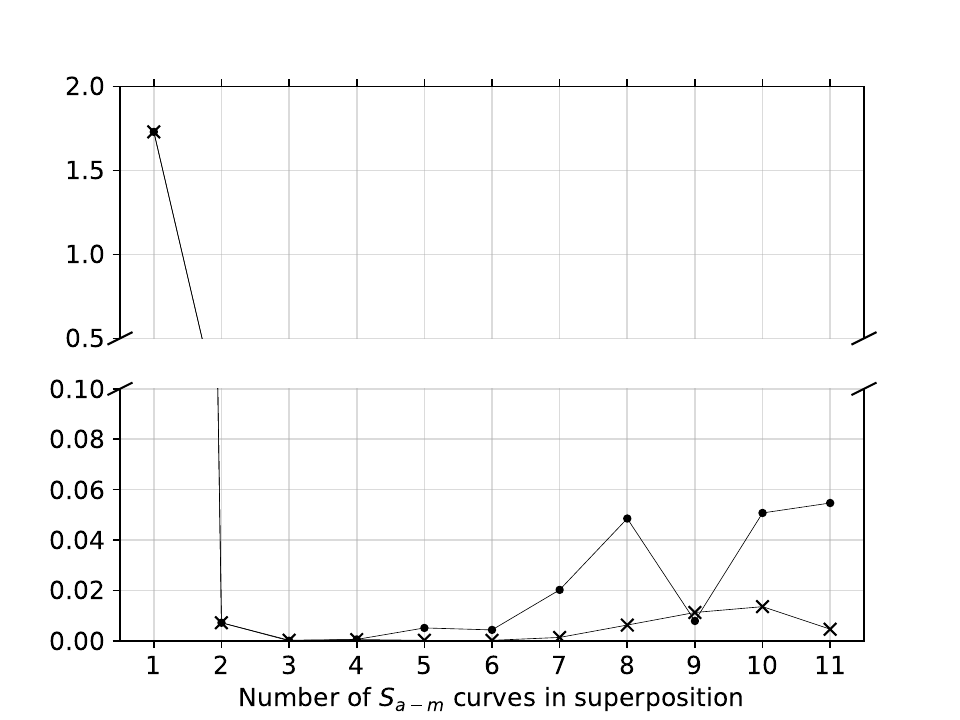}
		\caption{Estimation of $a$ stabilizes to some extent as $n$ increases.}
		\label{fig4a}
	\end{subfigure}
	\begin{subfigure}[t]{0.45\textwidth}
		\includegraphics[width=\textwidth]{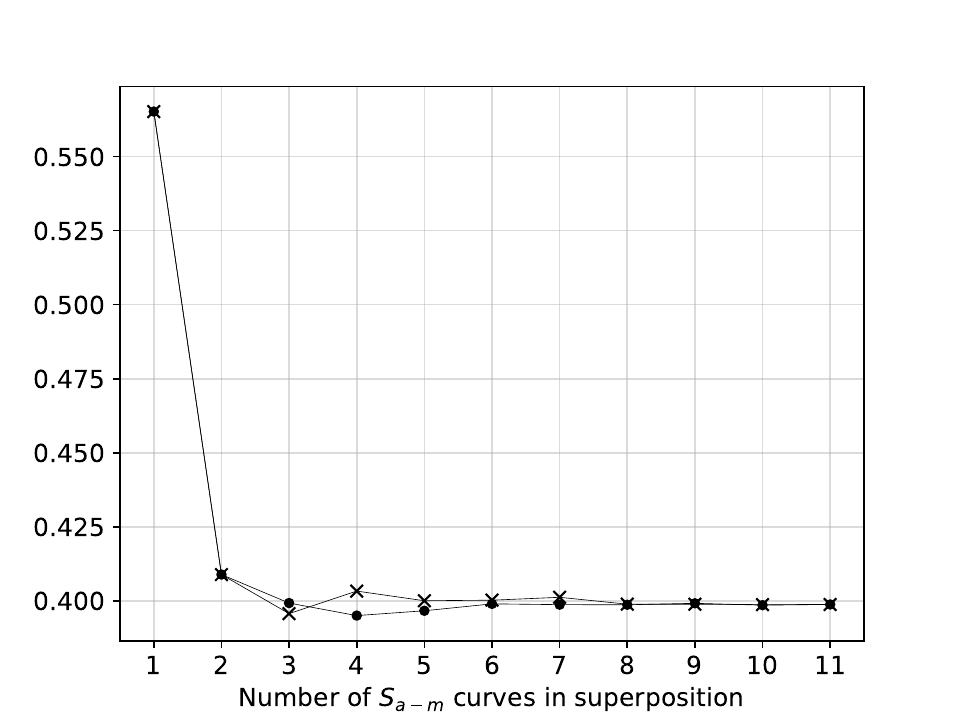}
		\caption{The maximum slope $m$ converge to a more accurate estimation as $n$ increases.}
		\label{fig4b}
	\end{subfigure}
	\caption{Variation of parameters $a$ and $m$ based on the fits on Gauss error function whose derivative is $\tilde{y}$ for two different initial conditions $p_i=1,\ m_i=1$ (shown as `$\times$') and $p_i=0,\ m_i=\text{slope at }x_{ci}\text{ between data points}$ (shown as `$\bullet$') with $n$ as the $x-$ axis.}
	\label{fig4}
\end{figure}

As shown in Fig. \ref{fig2}, $S_{\text{a-m}}$ superposition has fit the sigmoid and the bell-curves. Although there are many parameters involved we characterize the fits by choosing two representative parameters $a$ and $m$. Here, $a$ can be estimated directly from the fit. $m$ is the maximum slope value of the fitted curve or the peak value of the derived bell-curve, given by 
\begin{equation}
m=\left.\frac{dy_{\text{net}}}{dx}\right|_{\text{max}}.
\label{eq_m_max}
\end{equation}
In this section, we will fit the superposition on $S_{\text{exp}}$ for two different intervals of $x$ and on the Gauss error function for two different initial conditions and analyze the estimated $a$ and $m$. They are shown in Fig. \ref{fig3} and \ref{fig4}.
In Figs. \ref{fig3} and \ref{fig4}, unlike $a$, $m$ converges to a precise value irrespective of initial condition or the interval size as long as the maximum slope lies within the interval. The differences in $a$ are small when compared to unity. So, we can use the ratio $m/(1+a)$ as a measure to characterize a fit. This measure was introduced as an enzyme kinetic measure \cite{chem_kin} to analyze protease activity on synthetic peptide substrates. 

\begin{figure}[ht!]
	\centering
	\begin{subfigure}[t]{0.45\textwidth}
		\includegraphics[width=\textwidth]{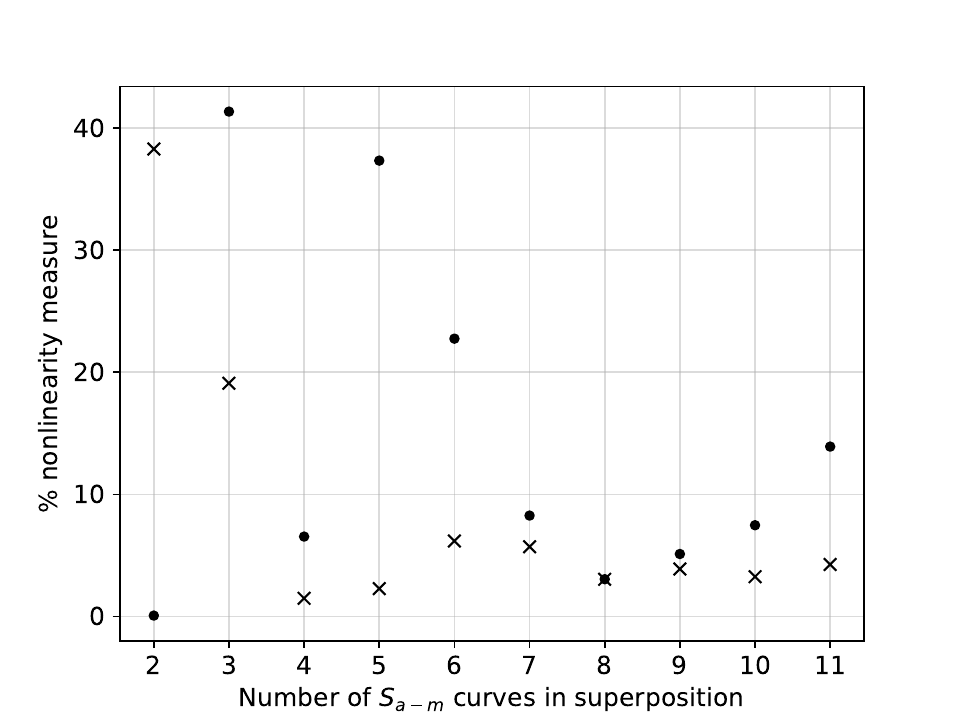}
		\caption{$S_{\text{exp}}$ with $x\in[-5,5]$(marker `$\bullet$') is more nonlinear than $S_{\text{exp}}$ with $x\in[-3,3]$ (marker `$\times$').}
		\label{fig5a}
	\end{subfigure}
	\begin{subfigure}[t]{0.45\textwidth}
		\includegraphics[width=\textwidth]{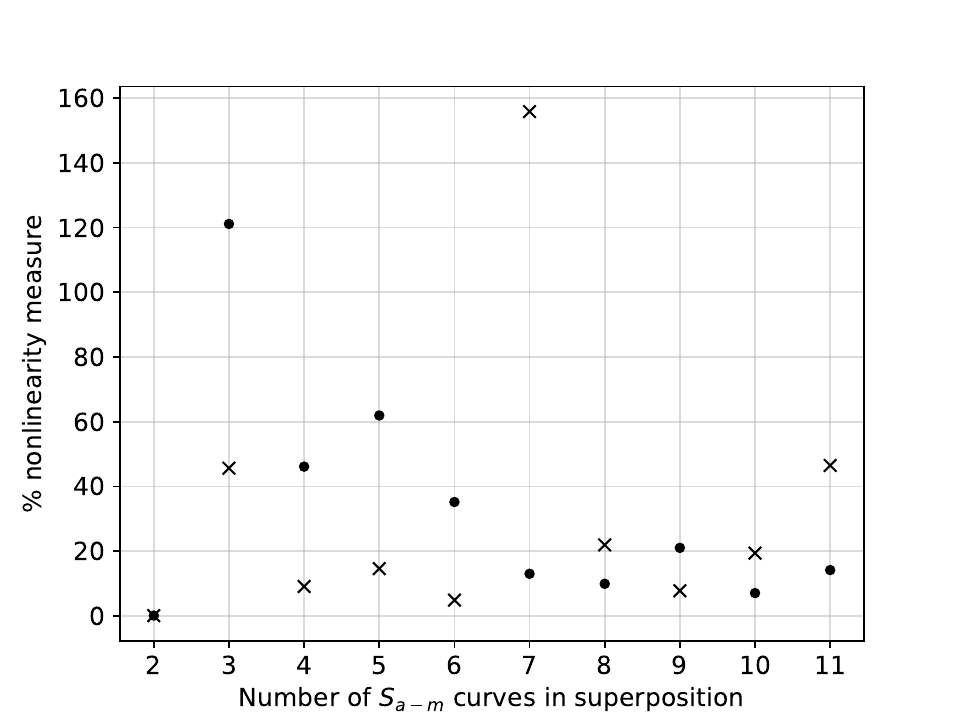}
		\caption{Initial conditions $p_i=1,\ m_i=1$ (shown as `$\times$') and $p_i=0,\ m_i=\text{slope at }x_{ci}\text{ between data points}$ (shown as `$\bullet$').}
		\label{fig5b}
	\end{subfigure}
	\begin{subfigure}[t]{0.45\textwidth}
		\includegraphics[width=\textwidth]{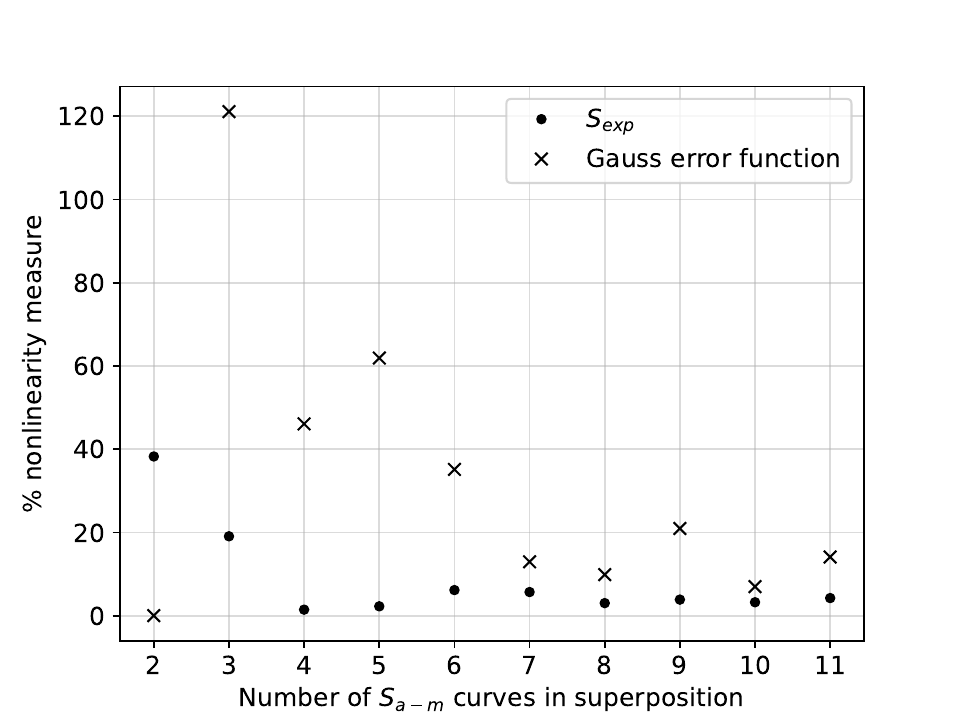}
		\caption{Gauss error function is more nonlinear than the sigmoid function within the interval $x\in[-3,3]$.}
		\label{fig5c}
	\end{subfigure}
	\caption{Percentage nonlinearity measures against the number of $S_{\text{a-m}}$ curves in superposition for (a) sigmoid function under different $x$ intervals (b) Gauss error function for different initial conditions (c) sigmoid and Gauss error function under the same initial conditions and $x$ interval.}
	\label{fig5}
\end{figure}
The other measure to characterize a fit is given by
\begin{equation}
\frac{\left|\sum_{i=1}^np_im_i-m\right|}{m}.
\label{eq_nonlin_meas}
\end{equation}
$\sum_{i=1}^np_im_i$ is $dy_\text{net}/dx$ with the nonlinearities turned off (keeping $a=0$), while $m$ depends on all the parameters including $a$. So, the above measure reflects the role of nonlinearities in the fit, hence this measure is referred to as percentage nonlinearity measure. As shown in Fig. (\ref{fig5a}), sigmoid function is more nonlinear on a wider interval $x\in[-5,5]$ than $x\in[-3,3]$ so the percentage nonlinearity of the former interval is higher than the later. However, the measure changes with the initial conditions of Gauss error fit shown in Fig. \ref{fig5b}. In Fig. \ref{fig5c}, the nonlinearity measure for sigmoid and Gauss error functions are compared for $x\in[-3,3]$ with same initial conditions. Gauss error function is observed to be more nonlinear than the sigmoid function in the same $x$ interval.

To summarize this section, the measures available are
\begin{enumerate}
\item The height of the bell- curve $m$. We can precisely obtain this measure as we increase $n$ without much dependence on initial conditions.
\item The ratio $m/(1+a)$. This measure works well as we increase $n$. However, uncertainty still exists due to fluctuations in $a$. Also. this measure is dependent on initial conditions.
\item Nonlinearity measure Eqn. (\ref{eq_nonlin_meas}). A curve is more nonlinear if it has a narrower peak. For example in Fig. \ref{fig2b}, sigmoid bell-curve is less nonlinear than the normal distribution function as suggested by Fig. \ref{fig5c}. 
\end{enumerate}
While $m$ is an absolute measure, the nonlinearity measure is a relative measure. The ratio $m/(1+a)$ is useful in ranking subsets of a dataset such as protease activity \cite{chem_kin}.
\section{Pattern Recognition}
\label{sec_iris}
As a nonlinear dynamic model the superposed form has fit biological growth data \cite{shruti_Ssum, shruti_vij2} and chemical kinetic data \cite{chem_kin}. In this section, we will use the $\sam$ curve and its superposed form as statistical models and fit the flower measurement data of iris plants. The dataset is publicly available in the UCI machine learning repository \cite{uci_iris}. This is a balanced dataset containing four attributes sepal length, sepal width, petal length and petal width of flowers belonging to plant species \textit{Iris setosa}, \textit{Iris versicolour} and \textit{Iris virginica}. There are 50 measurement values belonging to each species type. 

\begin{figure}[ht!]
	\centering
	\includegraphics[width=1.1\textwidth]{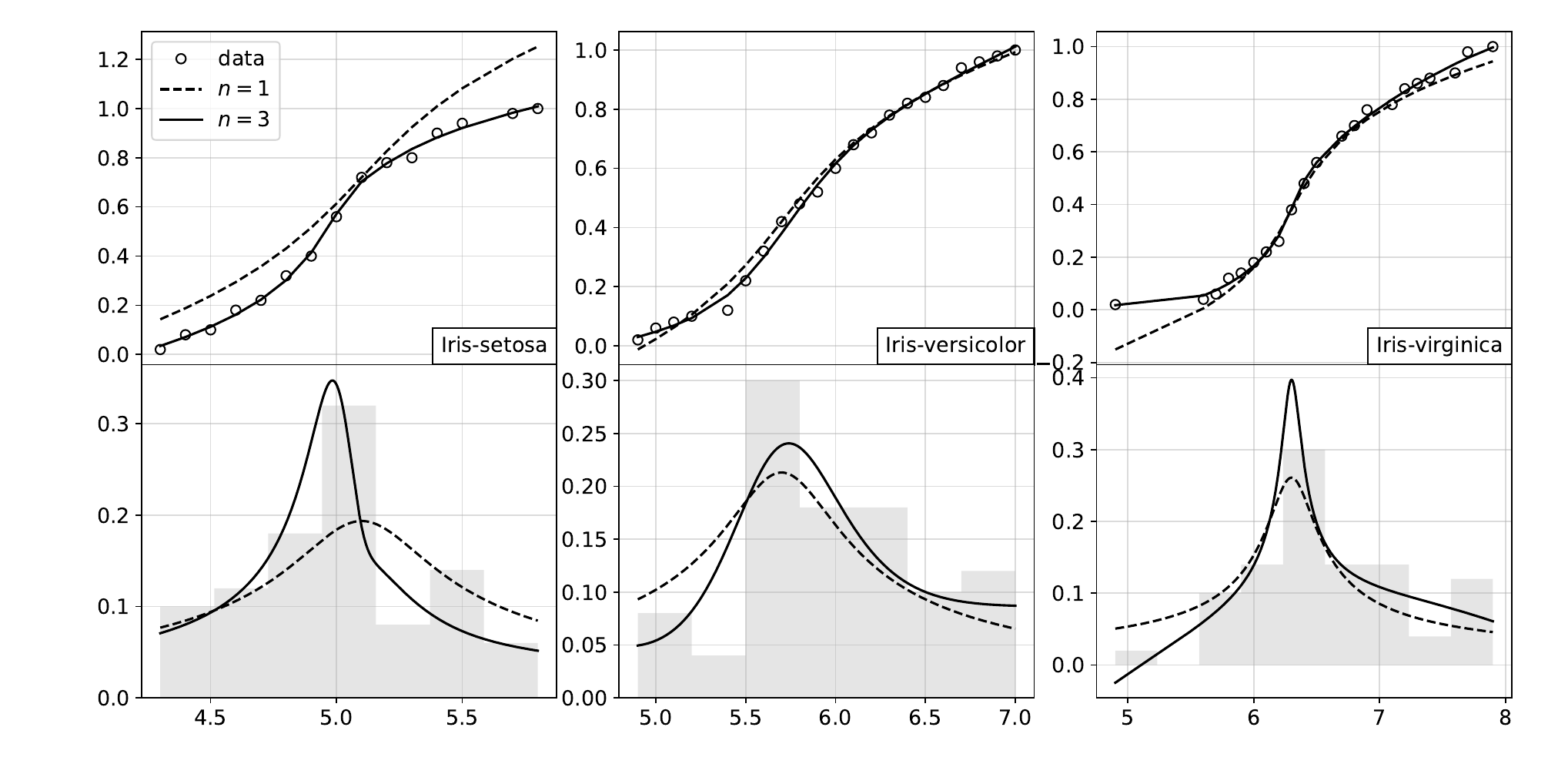}
	\caption{Fitted models on cumulative distributions of sepal length (in cm) of iris plants. The derived probability density curves are also shown below for each species. The parameter values for $n=1$ is shown in Table \ref{tab1_sepal_l} and for $n=3$ in Table \ref{tab2_sepal_l}}.
	\label{fig_sepal_length}
\end{figure}

\begin{table}[h!]
	\centering
	\begin{tabular}{|c|c|c|c|}
		\hline
	{Quantities} &{\textit{Iris setosa}}& {\textit{Iris versicolor}}&{\textit{Iris virginica}}\\\hline
	$a$ & 1.519780 & 2.295256 & 4.896959 \\\hline
	$m$ & 1.086830 & 0.772745 & 0.901669 \\\hline
	$x_c$ & 5.1cm & 5.7cm & 6.3cm \\\hline
	$y_c$ & 0.72 & 0.42 & 0.38 \\\hline
	$\bar{m}$ & 0.193602 & 0.213034 & 0.260928 \\\hline
	\end{tabular}
	\caption{Parameter values for $n=1$ sepal length distribution models in Fig. \ref{fig_sepal_length} with initial conditions $a=1$ and $m=0.1$, with the constraint $a>1e-9$.}
	\label{tab1_sepal_l}
\end{table}

\begin{table}[h!]
	\centering
	\begin{tabular}{|c|c|c|c|}
		\hline
		{Quantities} &{\textit{Iris setosa}}& {\textit{Iris versicolor}}&{\textit{Iris virginica}}\\\hline
		$a$ & 1.496536 & 0.008575 & 0.000256 \\\hline
		$p_1$ & -0.136528 & 0.607929 & -1001.161425 \\\hline
		$m_1$ & 0.757782 & 1.911407 & 0.461323 \\\hline
		$p_2$ & 0.808445 & -0.640305 & 712.868633 \\\hline
		$m_2$ & 2.722941 & 4.818308 & 0.648064 \\\hline
		$p_3$ & -0.171879 & 0.308483 & 0.003696 \\\hline
		$m_3$ & 4.248980 & 8.906210 & 295.540954 \\\hline
		$x_c$ & 5.4cm, 5cm,  5.1cm & 5.5cm, 5.6cm, 5.7cm & 6.4cm, 6.7cm, 6.3cm \\\hline
		$y_c$ & 0.9, 0.56, 0.72 & 0.22, 0.32, 0.42 & 0.48, 0.66, 0.38 \\\hline
		$m$ & 1.679906 & 0.844489 & 1.195249 \\\hline
		NL & 18.591877 & 2.399107 & 1.929964 \\\hline
		$\bar{m}$ & 0.347406 & 0.24059 & 0.397176 \\\hline
	\end{tabular}
	\caption{Parameter values for $n=3$ sepal length distribution models in Fig. \ref{fig_sepal_length} with initial condition $a=1,\ p_i=1$ and $m_i=1$ with the constraint $a>1e-9$. $\bar{m}$ represents probability density estimates. $m$ is obtained using Eqn. (\ref{eq_m_max})}
	\label{tab2_sepal_l}
\end{table}

\begin{figure}[ht!]
	\centering
	\includegraphics[width=1.1\textwidth]{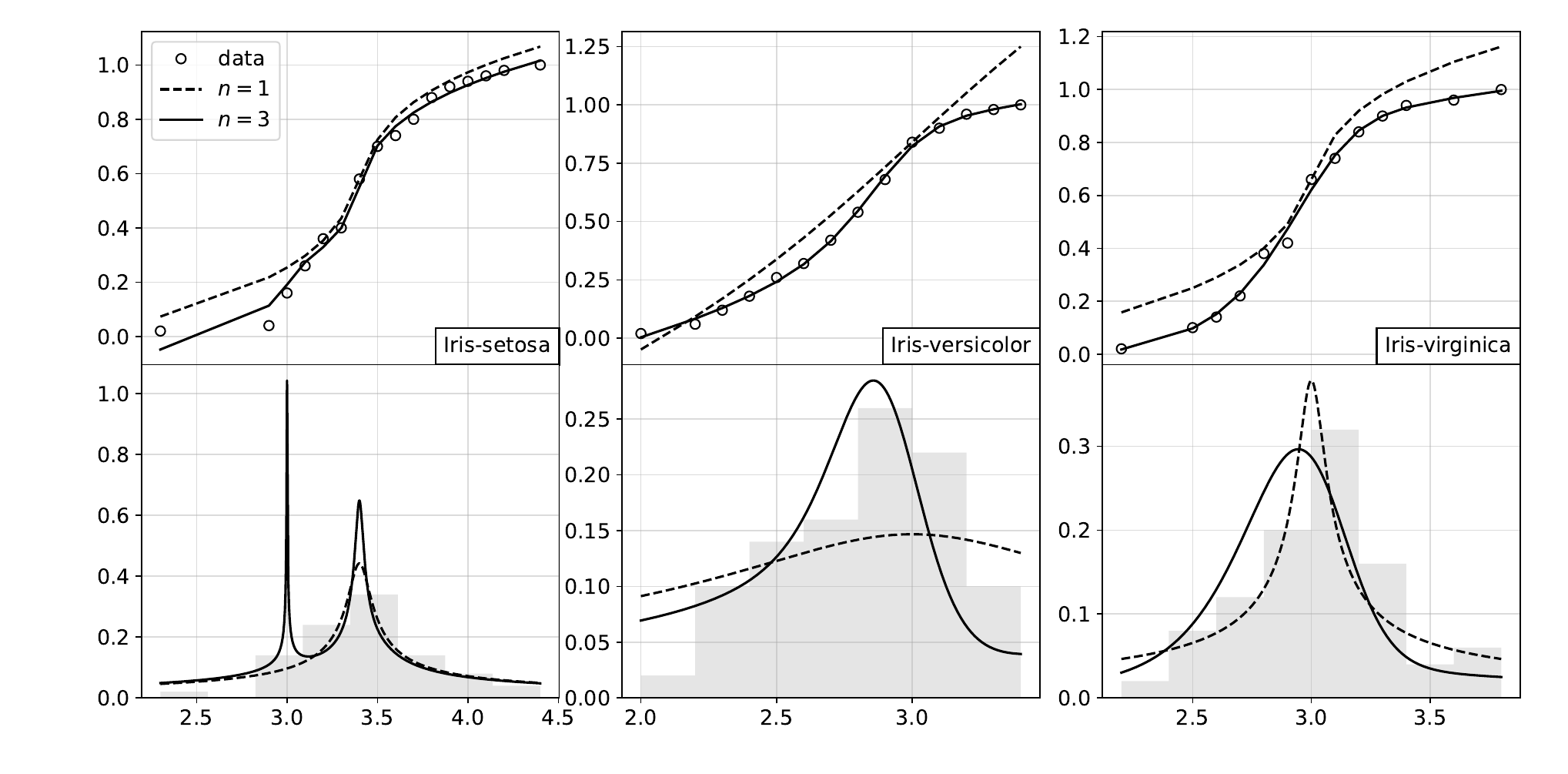}
	\caption{Fitted models on cumulative distributions of sepal width (in cm) of iris plants. The derived probability density curves are also shown below for each species. The parameter values for $n=1$ is shown in Table \ref{tab1_sepal_w} and for $n=3$ in Table \ref{tab2_sepal_w}}
	\label{fig_sepal_width}
\end{figure}

\begin{table}[h!]
	\centering
	\begin{tabular}{|c|c|c|c|}
		\hline
		{Quantities} &{\textit{Iris setosa}}& {\textit{Iris versicolor}}&{\textit{Iris virginica}}\\\hline
		$a$ & 11.216869 & 0.257162 & 9.462835 \\\hline
		$m_1$ & 1.789357 & 1.068846 & 2.127825 \\\hline
		$x_c$ & 3.4cm & 3.cm & 3.cm \\\hline
		$y_c$ & 0.58 & 0.84 & 0.66 \\\hline
		$\bar{m}$ & 0.441987 & 0.146775 & 0.378492 \\\hline
	\end{tabular}
	\caption{Parameter values for $n=1$ sepal width distribution models in Fig. \ref{fig_sepal_width} with initial condition $a=1$ and $m=0.1$ with the constraint $a>1e-9$.}
	\label{tab1_sepal_w}
\end{table}

\begin{table}[h!]
	\centering
	\begin{tabular}{|c|c|c|c|}
		\hline
		{Quantities} &{\textit{Iris setosa}}& {\textit{Iris versicolor}}&{\textit{Iris virginica}}\\\hline
		$a$ & 219.884500 & 225514.913420 & $1.185\times 10^{-7}$ \\\hline
		$p_1$ & 0.285454 & -4070.946274 & 0.000900 \\\hline
		$m_1$ & 13.519062 & 0.002175 & -3222.157056 \\\hline
		$p_2$ & -0.694127 & -9369.372072 & 0.001334 \\\hline
		$m_2$ & -0.191303 & -0.002235 & -1892.498683 \\\hline
		$p_3$ & -2.213338 & 4938.919811 & -0.002209 \\\hline
		$m_3$ & -1.086762 & -0.002283 & -2898.034947 \\\hline
		$x_c$ & 3cm, 3.5cm, 3.4cm & 2.8cm, 2.9cm, 3cm & 3.2cm, 2.8cm, 3cm \\\hline
		$y_c$ & 0.16, 0.7, 0.58 & 0.54, 0.68, 0.84 & 0.84, 0.38, 0.66 \\\hline
		$m$ & 3.491381 & 1.501607 & 1.488202 \\\hline
		NL & 83.229072 & 45.780925 & 34.307505 \\\hline
		$\bar{m}$ &  1.043143 & 0.284656 & 0.296565 \\\hline
	\end{tabular}
	\caption{Parameter values for $n=3$ sepal width distribution models in Fig. \ref{fig_sepal_width} with initial condition $a=1,\ p_i=1$ and $m_i=-1$ with the constraint $a>1e-9$. $\bar{m}$ represents probability density estimates. $m$ is obtained using Eqn. (\ref{eq_m_max})}
	\label{tab2_sepal_w}
\end{table}

\begin{figure}[ht!]
	\centering
	\includegraphics[width=1.1\textwidth]{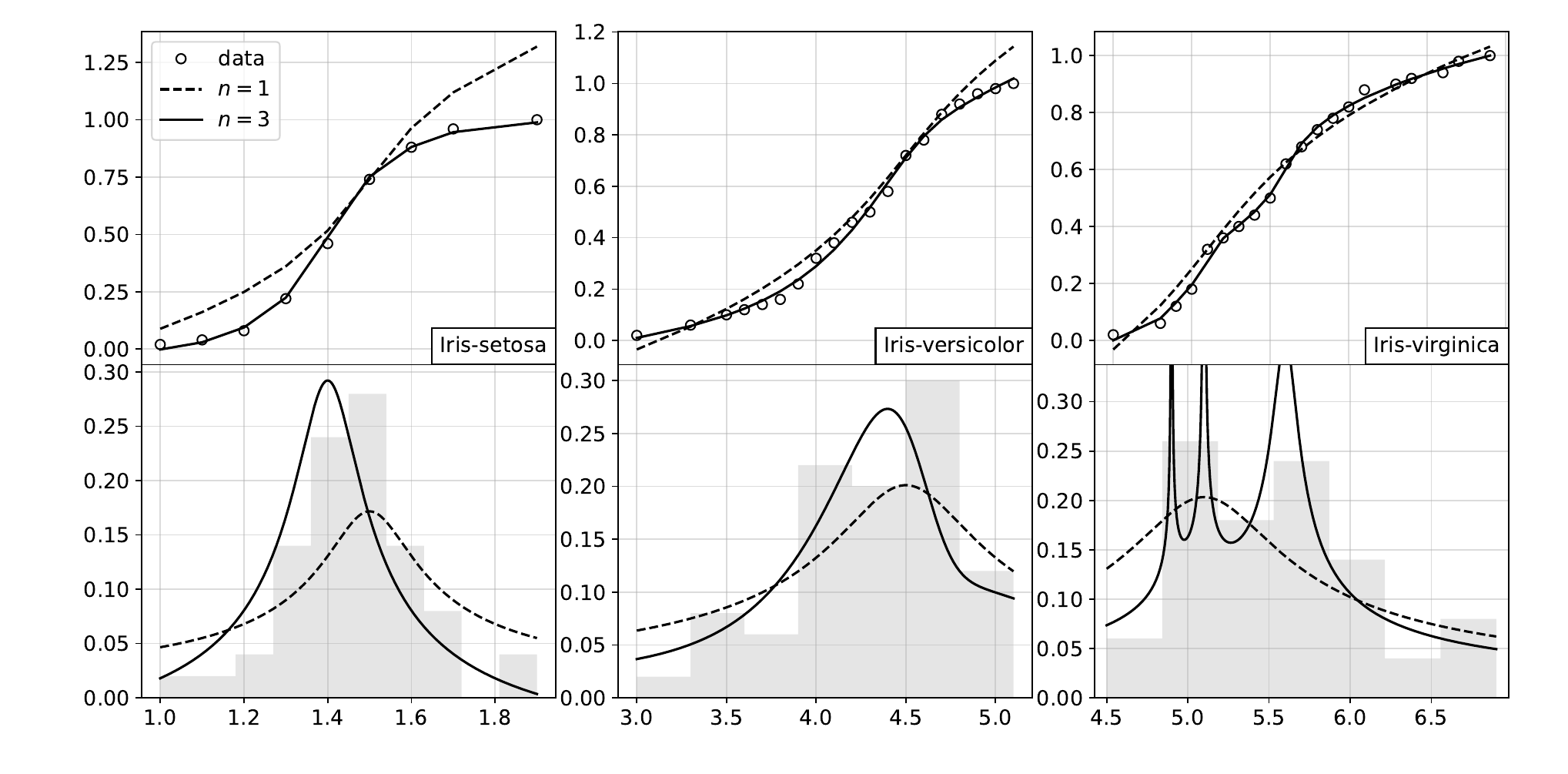}
	\caption{Fitted models on cumulative distributions of petal length (in cm) of iris plants. The derived probability density curves are also shown below for each species. The parameter values for $n=1$ is shown in Table \ref{tab1_petal_l} and for $n=3$ in Table \ref{tab2_petal_l}}
	\label{fig_petal_l}
\end{figure}

\begin{table}[h!]
	\centering
	\begin{tabular}{|c|c|c|c|}
		\hline
		{Quantities} &{\textit{Iris setosa}}& {\textit{Iris versicolor}}&{\textit{Iris virginica}}\\\hline
	$a$ & 2.109128 & 1.264663 & 1.495541 \\\hline
	$m_1$ & 2.470337 & 0.866053 & 0.694955 \\\hline
	$x_c$ & 1.5cm & 4.5cm & 5.1cm \\\hline
	$y_c$ & 0.74 & 0.72 & 0.32 \\\hline
	$\bar{m}$ & 0.17176 & 0.201067 & 0.203347 \\\hline
	\end{tabular}
	\caption{Parameter values for $n=1$ petal length distribution models in Fig. \ref{fig_petal_l} with initial condition $a=1$ and $m=0.1$ with the constraint $a>1e-9$.}
	\label{tab1_petal_l}
\end{table}

\begin{table}[h!]
	\centering
	\begin{tabular}{|c|c|c|c|}
		\hline
		{Quantities} &{\textit{Iris setosa}}& {\textit{Iris versicolor}}&{\textit{Iris virginica}}\\\hline
		$a$ & 0.142671 & 12457.557555 & 423.488765 \\\hline
		$p_1$ & -7.976545 & 149.711137 & -0.113996 \\\hline
		$m_1$ & 0.033544 & -0.003129 & -86.947406 \\\hline
		$p_2$ & -0.315143 & -98.120756 & -3.945396 \\\hline
		$m_2$ & -12.926374 & 0.011918 & -0.240541 \\\hline
		$p_3$ & 11.942307 & 277.277777 & 0.151272 \\\hline
		$m_3$ & -0.047720 & 0.008416 & 199.732348 \\\hline
		$x_c$ & 1.6cm, 1.4cm, 1.5cm & 4cm, 4.7cm, 4.5cm&4.9cm, 5.6cm, 5.1cm \\\hline
		$y_c$ & 0.88, 0.46, 0.74 & 0.32, 0.88, 0.72 & 0.12, 0.62, 0.32 \\\hline
		$m$ & 3.236169 & 0.981885 & 4.128101 \\\hline
		NL & 0.001117 & 29.128698 & 894.998545 \\\hline
		$\bar{m}$ & 0.292215 & 0.273217 & 9.193634 \\\hline
	\end{tabular}
	\caption{Parameter values for $n=3$ petal length distribution models in Fig. \ref{fig_petal_l} with initial condition $a=1,\ p_i=1$ and $m_i=-1$ with the constraint $a>1e-9$. $\bar{m}$ represents probability density estimates. $m$ is obtained using Eqn. (\ref{eq_m_max})}
	\label{tab2_petal_l}
\end{table}

\begin{figure}[ht!]
	\centering
	\includegraphics[width=1.1\textwidth]{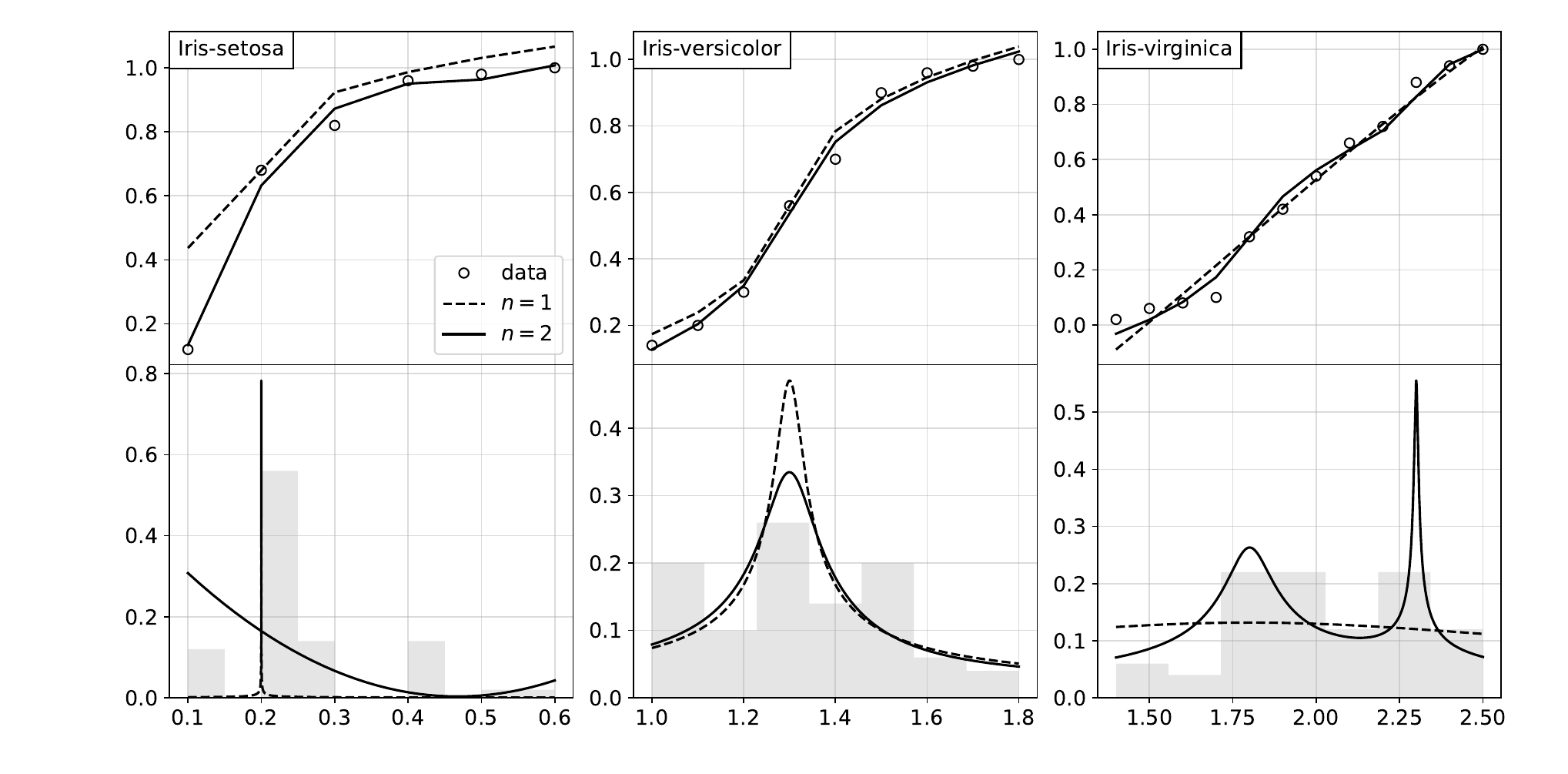}
	\caption{Fitted models on cumulative distributions of petal width (in cm) of iris plants. The derived probability density curves are also shown below for each species. The parameter values for $n=1$ is shown in Table \ref{tab1_petal_w} and for $n=3$ in Table \ref{tab2_petal_w}}
	\label{fig_petal_w}
\end{figure}

\begin{table}[h!]
	\centering
	\begin{tabular}{|c|c|c|c|}
		\hline
		{Quantities} &{\textit{Iris setosa}}& {\textit{Iris versicolor}}&{\textit{Iris virginica}}\\\hline
		$a$ & 3664.935912 & 11.989745 & 0.124025 \\\hline
		$m_1$ & 529.168336 & 3.591652 & 1.043516 \\\hline
		$x_c$ & 0.2cm & 1.3cm & 1.8cm \\\hline
		$y_c$ & 0.68 & 0.56 & 0.32 \\\hline
		$\bar{m}$ & 0.987226 &  0.470687  & 0.13179 \\\hline
	\end{tabular}
	\caption{Parameter values for $n=1$ petal width distribution models in Fig. \ref{fig_petal_w} with initial condition $a=1$ and $m=0.1$ with the constraint $a>1e-9$.}
	\label{tab1_petal_w}
\end{table}

\begin{table}[h!]
	\centering
	\begin{tabular}{|c|c|c|c|}
		\hline
		{Quantities} &{\textit{Iris setosa}}& {\textit{Iris versicolor}}&{\textit{Iris virginica}}\\\hline
		$a$ & 0.000334 & 458900.842161 & 1.38$\times 10^{-8}$  \\\hline
$p_1$ & -11142.446584 & 91.102275 & 0.000006 \\\hline
$m_1$ & 1.796412 & -0.001988 & 542267.203734 \\\hline
$p_2$ & 22184.395854 & 308.879028 & 0.000036 \\\hline
$m_2$ & 0.902318 & 0.009430 & 44873.328378 \\\hline
$x_c$ & 0.4cm, 0.2cm & 1.5cm, 1.3cm & 2.3cm, 1.8cm \\\hline
$y_c$ & 0.96, 0.68 & 0.9, 0.56 & 0.88, 0.32 \\\hline
$m$ & 6.599631 & 2.760746 & 3.629776 \\\hline
NL & 85.664366 & 1.058830 & 35.796550 \\\hline
$\bar{m}$ & 0.307843   & 0.334744   &    0.555527 \\\hline
	\end{tabular}
	\caption{Parameter values for $n=3$ petal width distribution models in Fig. \ref{fig_petal_w} with initial condition $a=1,\ p_i=1$ and $m_i=-1$ with the constraint $a>1e-9$.$\bar{m}$ represents probability density estimates. $m$ is obtained using Eqn. (\ref{eq_m_max})}
	\label{tab2_petal_w}
\end{table}

The models are fitted on the cumulative density function and the corresponding probability curves are obtained as the derivatives of fitted models. The cumulative density function is obtained as follows:
\begin{verbatim}
xax,cx = np.unique(np.sort(xdata),return_counts=True)
yax = np.cumsum(cx)/sum(cx)
\end{verbatim}
The {\tt (xax,yax)} plots and the fitted curves are shown on the top rows of Figs. \ref{fig_sepal_length} to \ref{fig_petal_w}. The models are given by
\begin{verbatim}
def sam(x,a,m):
  hatt = (-27*x*m/2/a+np.sqrt(729*(x*m/2/a)**2+27/a**3))
  S1 = -hatt**(1./3)/3
  S2 = hatt**(-1./3)/a
  return S1+S2
# Superposed sam curves
def sup_sam(params,xax,yax):
  #params is a dictionary of parameters a,m1,p1,m2,p2 and so on
  #xc,yc are lists of inflection point coordinates
  supmod = 0
  a = params['a'];xc = params['xc'];yc = params['yc']
  for i in range(1,len(xc)):
    supmod+=params['p'+str(i)]*(sam(xax-xc[i],a,params['m'+str(i)])+yc[i])
  return supmod
\end{verbatim}
We fit the above models using {\tt lmfit} package \cite{lmfit}. The corresponding histograms on the bottom rows of Figs. \ref{fig_sepal_length} to \ref{fig_petal_w} are obtained as follows:
\begin{verbatim}
import matplotlib.pyplot as plt
c,b = np.histogram(xdata,bins='auto',density=True)
plt.stairs(c/sum(c),b,fill=True,color='k',alpha=0.1);
\end{verbatim}
The derivatives of the models are obtained as follows.
\begin{verbatim}
def sam_der(x,a,m):
  return m/(1+3*a*sam(x,a,m)**2)
def sup_der(x,params):
  derval=0;
  a=params['a'];xc = params['xc'];yc = params['yc']
  for i in range(1,len(xc)):
    derval+=params['p'+str(i)]*params['m'+str(i)]/(1+3*a*sam(x-xc[i],a,params['m'+str(i)])**2)
  return derval
\end{verbatim}
In order to compare with the histograms, the derivatives of the models are normalized for $n=1$ as {\tt sam\_der(x,a,m)/np.sum(sam\_der(b,a,m))} and 
for $n>1$ as {\tt sup\_der(x,params)/np.sum(sup\_der(b,params))}. Therefore, the peak of a normalized bell-curve is denoted by $\bar{m}$. The inflection points are chosen to be the most frequently observed measurements ({\tt xax[np.argsort(cx)][::-1]}).
\begin{figure}[ht!]
	\centering
	\includegraphics[width=0.7\textwidth]{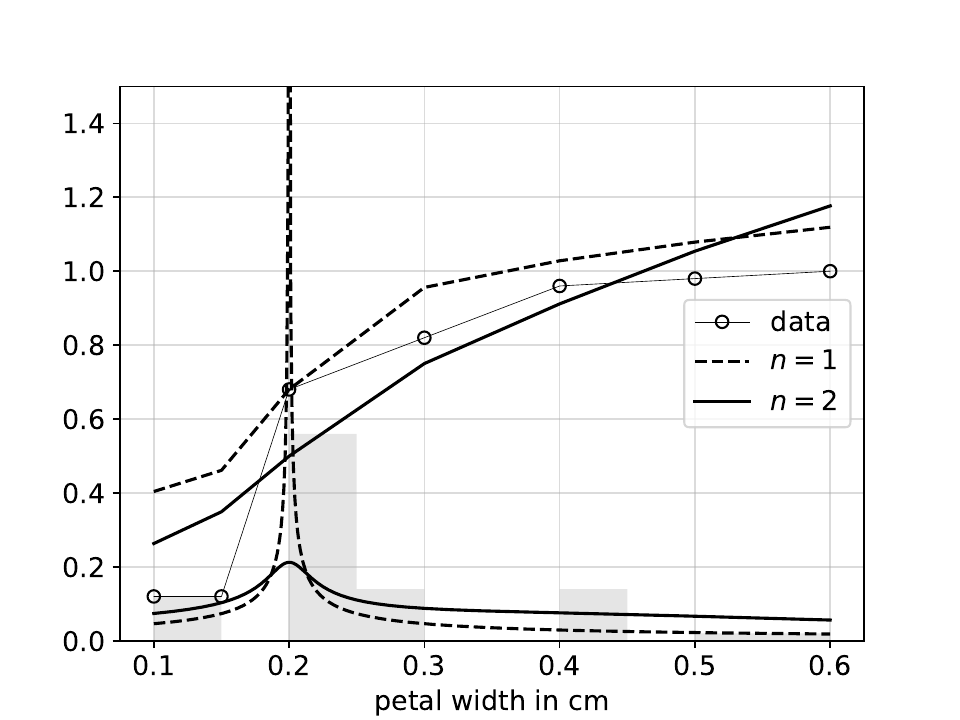}
	\caption{Fitted models on cumulative distributions of petal width (in cm) of Iris setosa after introducing a point at $x=0.15cm$.}
	\label{fig_petal_w_corr}
\end{figure}
\subsection{Results and discussions}
Estimated values of parameters of the fitted curves of Figs. \ref{fig_sepal_length} to \ref{fig_petal_w} are provided in Tables \ref{tab1_sepal_l} to \ref{tab2_petal_w}. The key discussions based on the fitted models are provided below:
\begin{enumerate}
\item \textbf{Sepal length:} For $n=1$ model, the $a$ values in Table \ref{tab1_sepal_l} reveal that \textit{Iris virginica} bell-curve is more centered and less distributed. This is inferred from the high $a$ values which results in a narrow peak as shown in Fig. \ref{fig1b}. However, when we consider $n=3$ model, the nonlinearity measure NL in Table \ref{tab2_sepal_l} is highest for \textit{Iris setosa} data. This may be because the bell-curves of \textit{Iris versicolor} and \textit{Iris virginica} data are more evenly distributed with wider $x-$ intervals. Hence, the NL values are less than \textit{Iris setosa} data which has a skewed bell-curve in Fig. \ref{fig_sepal_length}. \textit{Iris setosa} data has a very narrow range of sepal length with a peak at around 4.9cm. It can be noted that $n=1$ model does not fit well for \textit{Iris setosa} sepal length measurements for the chosen inflection point. However, $n=3$ reveals the actual inflection point and fits well for \textit{Iris setosa} values.

\item \textbf{Sepal width:} Considering the $a$ values of Table \ref{tab1_sepal_w}, it is observed that the sepal width measurements of \textit{Iris setosa} have a narrow peak centered around 3.4cm, however for \textit{Iris versicolor} the distribution is more uniform and it has a minimal $a$ value. From Fig. \ref{fig_sepal_width}, $n=3$ model is multimodal for \textit{Iris setosa} measurements and shows an extra peak at 3cm. However, $n=3$ model is unimodal for \textit{Iris versicolor} and \textit{Iris virginica} measurements. The NL values in Table \ref{tab2_sepal_w} suggests that the 
\textit{Iris versicolor} sepal width data has a narrower distribution than that of \textit{Iris virginica}. 

\item \textbf{Petal length:} In Fig. \ref{fig_petal_l}, $n=3$ model fits well for \textit{Iris setosa} data. The model is multimodal for the petal legnth data of \textit{Iris virginica}. The NL measure for \textit{Iris setosa} fit is far less than that of \textit{Iris versicolor}. This suggests that \textit{Iris setosa} data is more evenly distributed than \textit{Iris versicolor} data. $n=1$ model works well for data of \textit{Iris versicolor} and \textit{Iris virginica}.

\item \textbf{Petal width:} More than half of petal width data of \textit{Iris setosa} is 0.2cm. So, the distribution is highly centered around 0.2cm in Fig. \ref{fig_petal_w} resembling a degenerate distribution. However, $n=2$ model does not suggest a peak at 0.2cm. For \textit{Iris virginica} data $n=2$ model is multimodal. $n=1$ model suggests that petal width for \textit{Iris virginica} is distributed almost in a uniform manner. $n=2$ model does not provide us with the expected peak at 0.2cm for \textit{Iris setosa} data. A better fit for $n=2$ is obtained in Fig. \ref{fig_petal_w_corr} by introducing a point at x=0.15cm with zero frequency. Now, we do get a peak at 0.2cm in the bell-shaped curve of Fig. \ref{fig_petal_w_corr} corresponding to $n=2$.

\item From above points, the recognizable patterns are as follows:
\begin{itemize}
\item If sepal length is less than 5.5cm then sample is more likely an \textit{Iris setosa} plant. If it is more than 6cm, then \textit{Iris virginica} is the most likely category.
\item Sepal widths are similar in all the three categories of iris plants.
\item \textit{Iris setosa} is more distinguishable using petal length and petal width features.
\item If the petal length is around 5cm or more then \textit{Iris virginica} is the most likely category.
\end{itemize}
\end{enumerate}
The usefulness and drawbacks of $\sam$ model and its superposition are presented in Table \ref{tab_compar}.
\begin{table}[htbp]
	\centering
	\begin{tabular}{|p{0.5\linewidth}|p{0.5\linewidth}|}
		\hline
$n=1$ ($\sam$ curves) & $n>1$ (superposed $\sam$ form)\\\hline\hline
		2- parametric & 5- (or more) parametric\\\hline
		$m$ is independent of $a$ & $m$ depends on $a$\\\hline
Robust estimates for various initial conditions & Sensitive to initial conditions\\\hline
Sensitive to chosen point of inflection & Can be used to determine the point of inflection\\\hline
Fits well for less nonlinear distributions and may underfit data & May fit highly nonlinear data and can result in overfitting of data\\\hline
Applicable for unimodal distributions & Applicable for unimodal and multimodal distributions \\\hline
	\end{tabular}
	\caption{Comparison of $\sam$ model and its superposition.}
	\label{tab_compar}
\end{table}
\section{Conclusions}
In this work we have presented an algebraic approach to model sigmoidal and bell-shaped patterns in data. This approach involves introducing a singular perturbation to the straight line equation using a positive parameter. The perturbation introduces a nonlinear adjustment to the $y-$axis that keeps the straight lines bounded even for high slope values. The resulting curves, although nonlinear, are superposable with a common adjustment parameter. The superposed form can be used to fit highly nonlinear data of exponentially varying nature and can be used to quantify the nonlinearity of data. This ambitious attempt to use an algebraic approach has its drawbacks, such as introduction of more parameters and sensitivity to initial conditions. This approach has been used to model a classical dataset of various flower measurements of \textit{Iris} plant species and inferences have been made from the fitted model parameters.
	\bibliographystyle{unsrt}

\end{document}